\documentclass{article}

\PassOptionsToPackage{numbers,compress,sort}{natbib}
\usepackage[final,main]{neurips_2025}

\usepackage[utf8]{inputenc} %
\usepackage[T1]{fontenc}    %
\usepackage[hyphens]{url}   %
\usepackage{hyperref}       %
\usepackage{booktabs}       %
\usepackage{amsfonts}       %
\usepackage{nicefrac}       %
\usepackage{microtype}      %
\usepackage{xcolor}         %

\usepackage{enumitem}
\usepackage{amsthm}
\usepackage{amsmath}
\usepackage{amssymb}

\usepackage{graphicx}
	\setkeys{Gin}{width=\textwidth, height=\textheight, keepaspectratio}
\usepackage[skip=0pt]{subcaption}
\newcommand{\descr}[1]{\noindent\textbf{#1}}

\DeclareMathOperator{\E}{\mathbb{E}}

\title{Tight Auditing of Differential Privacy in\\MST and AIM}

\author{%
  Georgi Ganev$^{1,2}$ \quad Meenatchi Sundaram Muthu Selva Annamalai$^1$ \quad Bogdan Kulynych$^3$ \\
  $^1$UCL \quad $^2$SAS \quad $^3$Lausanne University Hospital \\
  \texttt{\href{georgi.ganev.16@ucl.ac.uk}{georgi.ganev.16@ucl.ac.uk}}\\
}

\begin{document}

\theoremstyle{plain}
\newtheorem{theorem}{Theorem}[section]
\newtheorem{proposition}[theorem]{Proposition}
\newtheorem{lemma}[theorem]{Lemma}
\newtheorem{corollary}[theorem]{Corollary}
\theoremstyle{definition}
\newtheorem{definition}[theorem]{Definition}
\newtheorem{assumption}[theorem]{Assumption}
\theoremstyle{remark}
\newtheorem{remark}[theorem]{Remark}

\maketitle

\begin{abstract}
State-of-the-art Differentially Private (DP) synthetic data generators such as MST and AIM are widely used, yet tightly auditing their privacy guarantees remains challenging.
We introduce a Gaussian Differential Privacy (GDP)-based auditing framework that measures privacy via the full false-positive/false-negative tradeoff.
Applied to MST and AIM under worst-case settings, our method provides the first tight audits in the strong-privacy regime.
For $(\epsilon,\delta)=(1,10^{-2})$, we obtain $\mu_{emp}\approx0.43$ vs. implied $\mu=0.45$, showing a small theory-practice gap.

Our code is publicly available: \url{https://github.com/sassoftware/dpmm}.
\end{abstract}

\section{Introduction}

Privacy-preserving synthetic data, supported by formal guarantees such as Differential Privacy~(DP)~\cite{dwork2006calibrating, dwork2014algorithmic}, offers a promising approach for sharing sensitive tabular data while protecting individual privacy.
Under this paradigm, a generative model is trained on the sensitive data with carefully calibrated noise so its outputs (e.g., model parameters/generated samples) are approximately indistinguishable with respect to the inclusion/exclusion of any individual record (typically formalized by the privacy parameters $(\epsilon,\delta)$).
Then, the trained model can serve as a proxy for the original data, enabling the release of either the model or synthetic datasets sampled from it.

Among the rich literature on DP generative models~\cite{jordon2022synthetic, cristofaro2024synthetic}, MST~\cite{mckenna2021winning} and AIM~\cite{mckenna2022aim} consistently demonstrate strong privacy-utility tradeoffs~\cite{tao2022benchmarking, ganev2024graphical, ganev2025importance, chen2025benchmarking}.
They have won a NIST competition~\cite{nist2018differential}, been integrated into popular libraries~\cite{opendp2021smartnoise, qian2023synthcity, mahiou2025dpmm}, and supported a UK census data release~\cite{ons2023synthesising}.
Both follow the select-measure-generate paradigm~\cite{mckenna2022simple, mckenna2021winning}: selecting low-order marginals, measuring them with Gaussian noise, and generating synthetic data preserving these marginals.

Given the importance and widespread use of MST and AIM, it is essential to empirically validate their DP guarantees.
This motivates DP auditing~\cite{annamalai2026hitchhiker}, which quantifies privacy leakage by framing evaluation as a distinguishing game aligned with the DP definition, often realized via Membership Inference Attacks (MIAs)~\cite{shokri2017membership, hayes2019logan} (which assess the influence of a target record on the model).
The resulting leakage estimates are then compared to the claimed DP parameters to verify implementation correctness~\cite{nasr2021adversary, nasr2023tight, steinke2023privacy, annamalai2024you} or reveal potential violations~\cite{haney2022precision, casacuberta2022widespread, lokna2023group, annamalai2024you, ganev2025the}.

Although prior work has investigated DP auditing for MST~\cite{annamalai2024you, mahiou2025dpmm}, important gaps remain.
First, existing audits provide limited insight in the strong-privacy regime (e.g., $\epsilon=1$), yielding loose/null results unless the implementation exhibits issues such as data domain leakage or floating-point vulnerabilities.
Second, prior studies report results for a single $(\epsilon,\delta)$ setting rather than characterizing the full privacy tradeoff, which can lead to incomplete or even misleading conclusions~\cite{gomez2026gaussian}.

In this paper, we close these gaps by developing a Gaussian DP (GDP)~\cite{dong2022gaussian} auditing framework for DP synthetic data generators.
Building on prior empirical audits~\cite{annamalai2024you}, we adopt the GDP view, which summarizes privacy via a single parameter $\mu$ capturing the full false-positive/negative tradeoff (FPR/FNR) of a strong-adversary MIA.
We apply this framework to MST/AIM in a worst-case setup.

\begin{figure*}[t!]
    \vspace{-0.5cm}
	\centering
    \includegraphics[width=0.5\textwidth]{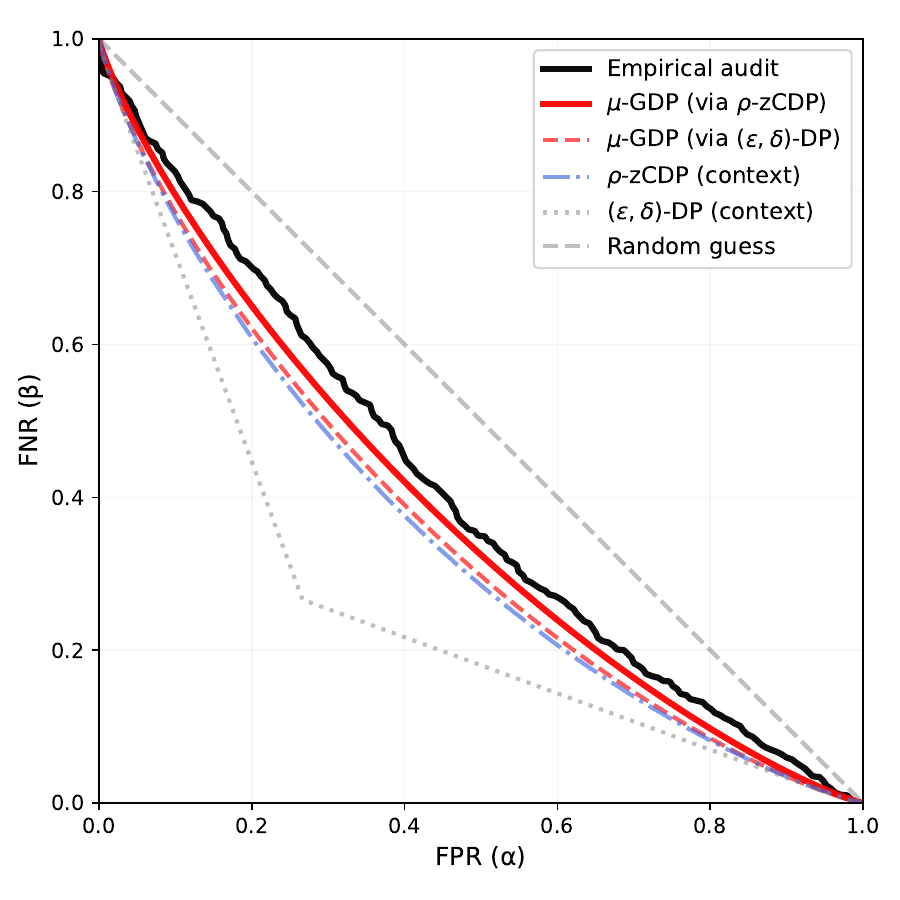}
    \vspace{-0.3cm}
	\caption{Empirical privacy tradeoff of MIA for MST/AIM compared to theoretical bounds under multiple DP notions.
    We report the full FPR--FNR tradeoff curve in the $\mu$-GDP framework, enabling direct comparison between empirical leakage and theory.
    At $(\epsilon,\delta)=(1,10^{-2})$, the empirical curve closely follows the implied $\mu$-GDP bound derived via $\rho$-zCDP (MST/AIM's accounting path), with $\mu_{emp}\approx0.43$ vs.~$\mu=0.45$.}
	\label{fig:tradeoff}
    \vspace{-0.5cm}
\end{figure*}

\descr{Main Contributions:}
\begin{itemize}[leftmargin=1.25em, itemsep=0.5pt, topsep=0.5pt]
    \item We present a GDP-centric auditing methodology that reports the full FPR--FNR tradeoff and places multiple DP notions on a common curve, rather than relying on a single $(\epsilon,\delta)$ value (see Fig.~\ref{fig:tradeoff}).
    \item We provide the first systematic and tight empirical audit of MST and AIM in the strong-privacy regime.
    For $(\epsilon,\delta)=(1,10^{-2})$, corresponding to an implied $\mu=0.45$, we obtain an empirical lower bound of $\mu_{emp}\approx0.43$, showing a small gap between theory and practice.
    \item We release our code to make GDP-based auditing reproducible and accessible to the community.
\end{itemize}

\section{DP Preliminaries}

Differential Privacy (DP)~\cite{dwork2006calibrating, dwork2014algorithmic} ensures that a randomized mechanism's output is insensitive to the inclusion/exclusion of any single record.
We review the DP notions relevant to this work.

\descr{$(\epsilon,\delta)$-DP.}
Approximate DP~\cite{dwork2014algorithmic} relaxes pure DP by allowing a failure probability $\delta$.

\begin{definition}[$(\epsilon,\delta)$-DP~\cite{dwork2014algorithmic}]
A randomized mechanism $\mathcal{M}:\mathcal{X}\to\mathcal{R}$ is $(\epsilon,\delta)$-DP if for all neighboring datasets $x,x'\in\mathcal{X}$ and measurable sets $S\subseteq\mathcal{R}$,
\[
\Pr[\mathcal{M}(x)\in S]\le e^\epsilon\Pr[\mathcal{M}(x')\in S]+\delta.
\]
\end{definition}

\descr{$\rho$-zCDP.}
Zero-concentrated DP (zCDP)~\cite{bun2016concentrated} characterizes privacy via R{\'e}nyi divergence~\cite{renyi1961measures} ($D_\alpha$ of order $\alpha$) and enables tight analysis of Gaussian mechanisms.

\begin{definition}[$\rho$-zCDP~\cite{bun2016concentrated}]
$\mathcal{M}$ satisfies $\rho$-zCDP if for all neighboring $x,x'$ and all $\alpha>1$,
\[
D_\alpha(\mathcal{M}(x)\|\mathcal{M}(x'))\le \rho\alpha,
\]
\end{definition}

\begin{theorem}[zCDP $\Rightarrow$ DP~\cite{bun2016concentrated}]
If $\mathcal{M}$ is $\rho$-zCDP, then for any $\delta>0$ it is
$(\rho+2\sqrt{\rho\log(1/\delta)},\,\delta)\text{-DP}$.
\end{theorem}

\descr{$\mu$-GDP.}
Under the hypothesis-testing view of DP~\cite[see, e.g.,][]{dong2022gaussian}, any DP mechanism limits the success of Membership Inference Attacks (MIAs) in the strong-adversary model, where the adversary knows $\mathcal{M}$ and two neighboring datasets $x,x'$ but not which was used.
An MIA is defined by a hypothesis test $\phi: \mathrm{dom}(\mathcal{M}) \to [0, 1]$, with values close to 1 indicating that dataset $x'$ was used, and vice versa.
For an MIA, we define $\beta = 1 - \E_{M(x)}[\phi]$ (FNR) and $\alpha = \E_{M(x')}[\phi]$ (FPR).
The tradeoff function $T(M(x), M(x'))(\alpha)$ gives the minimal achievable FNR $\beta^*$ at FPR level $\alpha$ (by Neyman-Pearson lemma~\cite{dong2022gaussian}).
Importantly for audits, any MIA must satisfy $\beta \ge \beta^*$.

Gaussian DP (GDP)~\cite{dong2022gaussian} characterizes privacy by lower bounding tradeoffs with a Gaussian curve:

\begin{definition}[$\mu$-GDP~\cite{dong2022gaussian}]
A mechanism $\mathcal{M}$ is $\mu$-GDP if for all neighboring $x,x'$,
\[
T(\mathcal{M}(x),\mathcal{M}(x'))(\alpha)\ge G_\mu(\alpha)\quad\forall \alpha\in[0,1],
\]
where and $G_\mu=T(\mathcal{N}(0,1),\mathcal{N}(\mu,1))$.
\end{definition}

\begin{theorem}[zCDP $\Rightarrow$ GDP~\cite{dong2022gaussian}]
If $\mathcal{M}$ is a Gaussian mechanism, or a composition of Gaussian mechanisms, satisfying $\rho$-zCDP, then it satisfies $\mu$-GDP with $\mu=\sqrt{2\rho}$.
\end{theorem}

\begin{theorem}[GDP $\Leftrightarrow$ DP~\cite{dong2022gaussian}]
$\mathcal{M}$ is $\mu$-GDP iff it is $(\epsilon,\delta(\epsilon))$-DP for all $\epsilon\ge0$, where
\[
\delta(\epsilon)=\Phi\!\left(-\frac{\epsilon}{\mu}+\frac{\mu}{2}\right)
- e^\epsilon\Phi\!\left(-\frac{\epsilon}{\mu}-\frac{\mu}{2}\right).
\]
\end{theorem}

\descr{MST and AIM}
follow the select-measure-generate paradigm~\cite{mckenna2021winning, mckenna2022simple}.
They use the Exponential mechanism~\cite{dwork2006our} to select informative marginals and the Gaussian mechanism~\cite{mcsherry2007mechanism} to measure counts.
Both start from one-way marginals; MST selects pairwise marginals forming a maximum spanning tree, while AIM iteratively selects higher-order marginals based on estimated importance.
Synthetic datasets are generated to match the noisy marginals via a graphical-model inference~\cite{mckenna2019graphical}.

Although privacy is specified as $(\epsilon,\delta)$-DP, both models internally convert this budget to a $\rho$-zCDP guarantee using the \citet{canonne2020discrete} conversion, selecting the smallest $\rho$ that implies the target $(\epsilon,\delta)$ guarantee.
Consistent with existing implementations~\cite{mckenna2019private, mahiou2025dpmm}, we adopt this numerically stable (though suboptimal~\cite{asoodeh2021three, riess2026optimal}) conversion in our experiments.

\section{GDP Auditing Framework of MST and AIM}

We adopt the empirical DP auditing framework of~\citet{annamalai2024you}, which views privacy auditing as a hypothesis test via a Membership Inference Attack~(MIA).
In short, a challenger trains many independent instances of a randomized mechanism on two neighboring datasets and provides their outputs to an adversary, who attempts to distinguish which dataset was used.
The resulting false-positive and false-negative rates define a point on the $(\alpha,\beta)$ tradeoff curve, which can be compared to theoretical DP frontiers and converted to an empirical $\mu$ under $\mu$-GDP.
Next, we describe the components of the framework and our specific instantiation.

\descr{MST and AIM.}
We study MST and AIM under a restricted configuration where only one-way marginals are measured.
The dependency graph is fixed and no higher-order marginals are selected, so MST and AIM reduce to the same independent-marginal model (we relax this assumption in App.~\ref{app:marginals}).
The full budget is allocated to these measurements, reducing the model to (Gaussian) noisy one-way marginals, which enables analysis in the $\mu$-GDP framework.
Also, we disable domain compression to avoid additional sources of bias~\cite{ganev2022robin} and use the implementation in the \emph{dpmm} library~\cite{mahiou2025dpmm}, which, to our knowledge, has no known DP vulnerabilities/violations.

\descr{Privacy Parameterization.}
As mentioned, both models convert the input $(\epsilon,\delta)$-DP budget to $\rho$-zCDP.
Because only Gaussian mechanisms are used, this $\rho$ directly maps to a $\mu$-GDP guarantee, which we treat as the theoretical target in our audits (we refer to it as implied $\mu$).

\descr{Neighboring Datasets and Target Record.}
We use a worst-case construction: $D_{out}$ contains 10 identical records $[0,0,0]$, and $D_{in}$ adds one target record $[1,1,1]$.
This maximizes the target's influence on all marginals and yields a stringent privacy test.
These worst-case settings are consistent with the DP definition and are often needed to obtain tight audits~\cite{nasr2021adversary, nasr2023tight, annamalai2024you}.

\descr{Threat Model.}
We consider a hybrid black/white-box adversary.
Black-box features are derived from synthetic data by evaluating all possible queries over the discrete domain, following the Querybased MIA~\cite{houssiau2022tapas}, while white-box features consist of the model's internal noisy one-way marginal counts.

\descr{DP Distinguishing Game.}
We train 10,000 independent models (5,000 per $D_{out}$/$D_{in}$) with $(\epsilon,\delta)=(1,10^{-2})$.
Each model releases a synthetic dataset of size 50 and noisy marginals.
The adversary receives balanced labeled outputs and splits them into 4,000 train, 2,000 validation, and 4,000 test samples.
They train an XGBoost classifier~\cite{chen2016xgboost} and select a decision threshold $\tau^*$ on validation to maximize their advantage (defined by TPR$-$FPR).
The test set is used only for final evaluation.

\descr{Statistical Estimation of $\mu$.}
Based on the adversary's predictions on $D_{test}$, we compute the TPR and FPR.
We then use a joint Bayesian approach inspired by~\citet{zanella2023bayesian} and derive a lower bound $\mu_{emp}$ such that the corresponding $\mu$-GDP region contains 90\% posterior mass.

This should be interpreted as a Bayesian lower bound on privacy.
As noted by~\citet{nasr2023tight}, such bounds do not necessarily yield valid frequentist coverage, though they are often empirically tighter.

\section{Experimental Results}
In this section, we present the experimental evaluation of our GDP auditing framework on MST/AIM.

\descr{Full FPR--FNR Tradeoff Curve.}
In Fig.~\ref{fig:tradeoff}, we show the empirical FPR--FNR tradeoff of our attack classifier across decision thresholds on the validation data, together with theoretical frontiers under multiple DP notions.
Expressing all guarantees in the $\mu$-GDP framework enables a direct comparison between empirical leakage and theory.
We observe that the empirical curve (solid black line) closely tracks the $\mu$ implied by the mechanism's internal $\rho$-zCDP accounting (solid red line), indicating that the implementation is consistent (and tight) with its privacy analysis.

In contrast, the direct $(\epsilon,\delta)\!\to\!\mu$ conversion (dashed red line) yields a larger $\mu$ (weaker privacy) than the $\mu$ implied by the $(\epsilon,\delta)\!\to\!\rho\!\to\!\mu$ accounting path used internally by MST/AIM.
This gap arises because the intermediate zCDP step relies on a conservative (though numerically stable) $(\epsilon,\delta)\!\to\!\rho$ conversion~\cite{canonne2020discrete}.
As a result, it enforces slightly stronger privacy (smaller $\mu$) than the direct GDP interpretation, which may reduce the achievable utility for the same input $(\epsilon,\delta)$.

\begin{figure*}[t!]
    \vspace{-0.5cm}
	\begin{minipage}[t]{0.485\linewidth}
        \centering
		\includegraphics[width=0.99\linewidth]{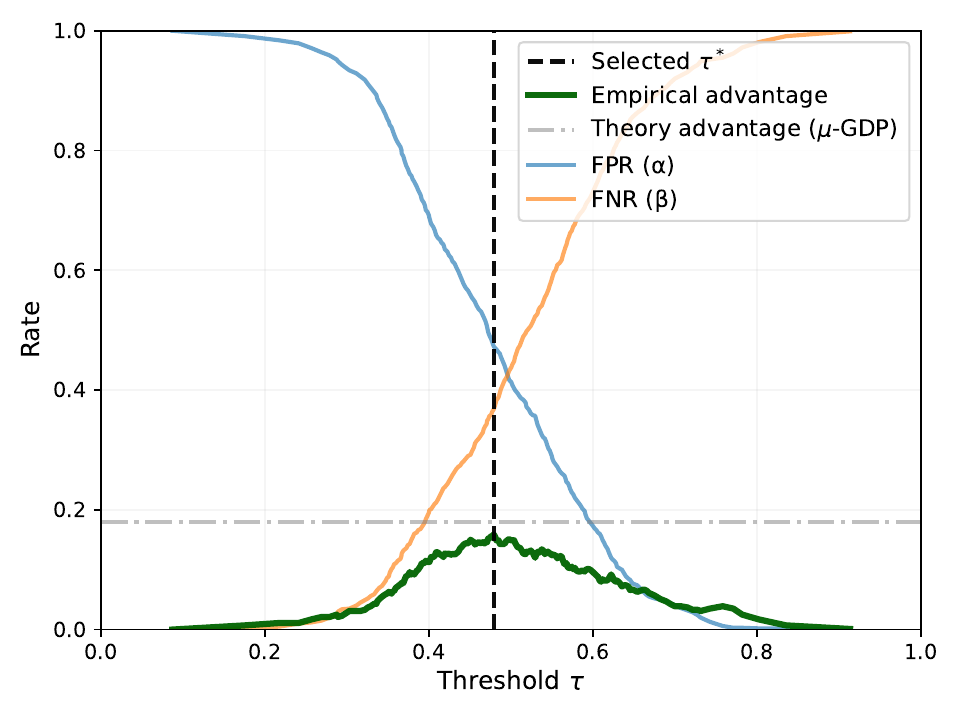}
        \vspace{-0.7cm}
		\caption{Threshold selection on validation data.
        FPR, FNR, and empirical advantage (TPR$-$FPR) as functions of the decision threshold.
        The selected threshold $\tau^*$ ($\approx0.48$) maximizes validation advantage.}
		\label{fig:valid}
	\end{minipage}
	\hfill
	\begin{minipage}[t]{0.485\linewidth}
        \centering
		\includegraphics[width=0.99\linewidth]{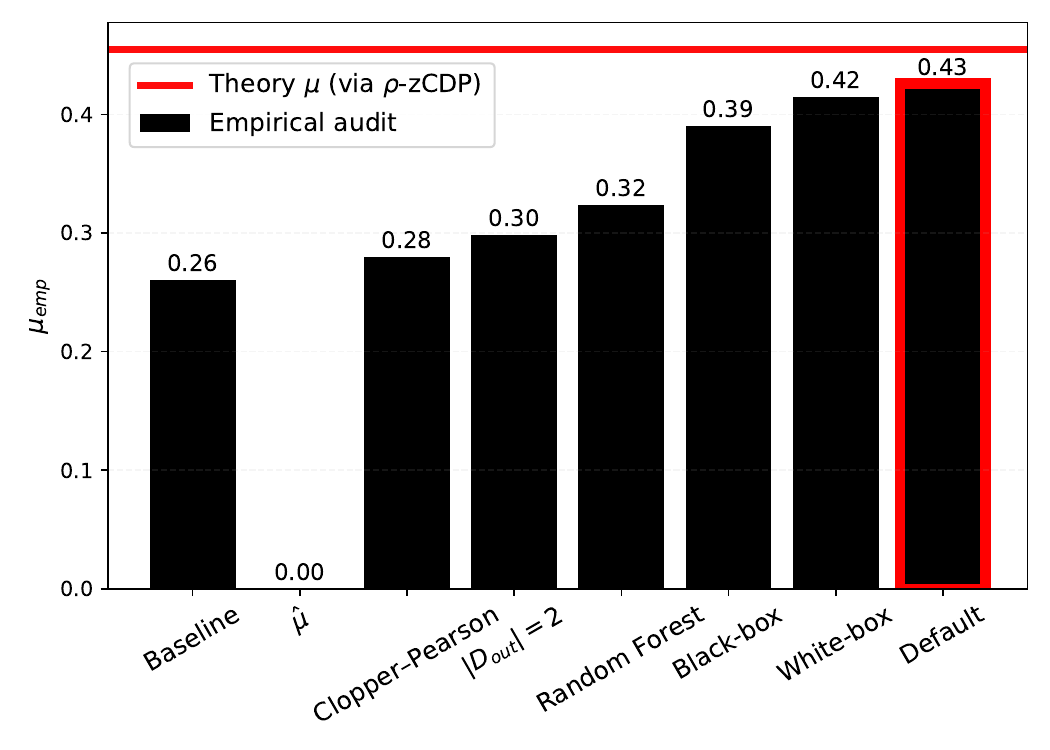}
        \vspace{-0.7cm}
		\caption{Ablation of audit choices and their effect on empirical privacy $\mu_{emp}$ on test data.
        ``Default'' is the main configuration; other bars modify one component at a time.
        Red line is the implied $\mu$ derived via $\rho$-zCDP.}
		\label{fig:abl}
    \end{minipage}
    \vspace{-0.5cm}
\end{figure*}

\descr{Tight Audit.}
Next, in Fig.~\ref{fig:valid}, we select the optimal decision threshold on validation data by maximizing empirical advantage, which exhibits smooth and stable behavior across thresholds.
Applying this fixed threshold to the held-out test data yields a tight lower-bound estimate (see Fig.~\ref{fig:abl}, ``Default'' configuration), with $\mu_{emp}\approx 0.43$ close to the implied $\mu=0.45$.

This establishes, to the best of our knowledge, the first tight audit of MST/AIM in the strong-privacy regime ($\epsilon=1$), as prior work reports zero empirical estimates~\cite{annamalai2024you, mahiou2025dpmm}.
A small theory-practice gap remains, but it is substantially narrower than previously observed.

\descr{Ablation Study.}
Finally, in Fig.~\ref{fig:abl}, we report an ablation study on the test data, where we vary one audit component at a time: threshold selection criterion, lower-bound estimation, $D_{out}$ size, classifier choice, and threat model.
We also include a naive white-box baseline that estimates $\mu$ from single noisy marginal.
From labeled in/out runs of this marginal, we compute a pooled variance and apply a $\chi^2$ confidence bound to derive a lower bound on $\mu = \Delta/\sigma$ (with $\Delta=1$) under Gaussian noise.
The ``Default'' configuration consistently yields the strongest (tightest) empirical lower-bound estimation.
Additionally, in App.~\ref{app:marginals}, we audit MST/AIM trained with higher-order marginals.

Notably, unstable threshold selection criteria (e.g., maximizing the estimated $\hat{\mu}$ on validation data) can select overly small thresholds, leading to excessive FPR and collapsing the empirical lower bound to zero.
This explains why prior work reports $\epsilon_{emp}=0$ results in the strong-privacy regime~\cite{annamalai2024you, mahiou2025dpmm}.

\section{Conclusion}

In this paper, we introduced a GDP-based auditing framework for MST and AIM that measures privacy through the full false-positive/false-negative tradeoff, enabling the first tight empirical audit in the strong-privacy regime.
Beyond these two models, our framework readily extends to other DP generative approaches based on the Gaussian mechanism.
We hope that our work (and the accompanying open-source implementation) will encourage researchers and practitioners to systematically validate the privacy guarantees of newly proposed and deployed synthetic data models.

\setlength{\bibsep}{3.5pt plus 15ex}
%

\bibliographystyle{plainnat}

\appendix

\section{Auditing MST/AIM Trained with Higher-Order Marginals}
\label{app:marginals}

\begin{figure*}[t!]
    \vspace{-0.5cm}
	\centering
    \includegraphics[width=0.75\textwidth]{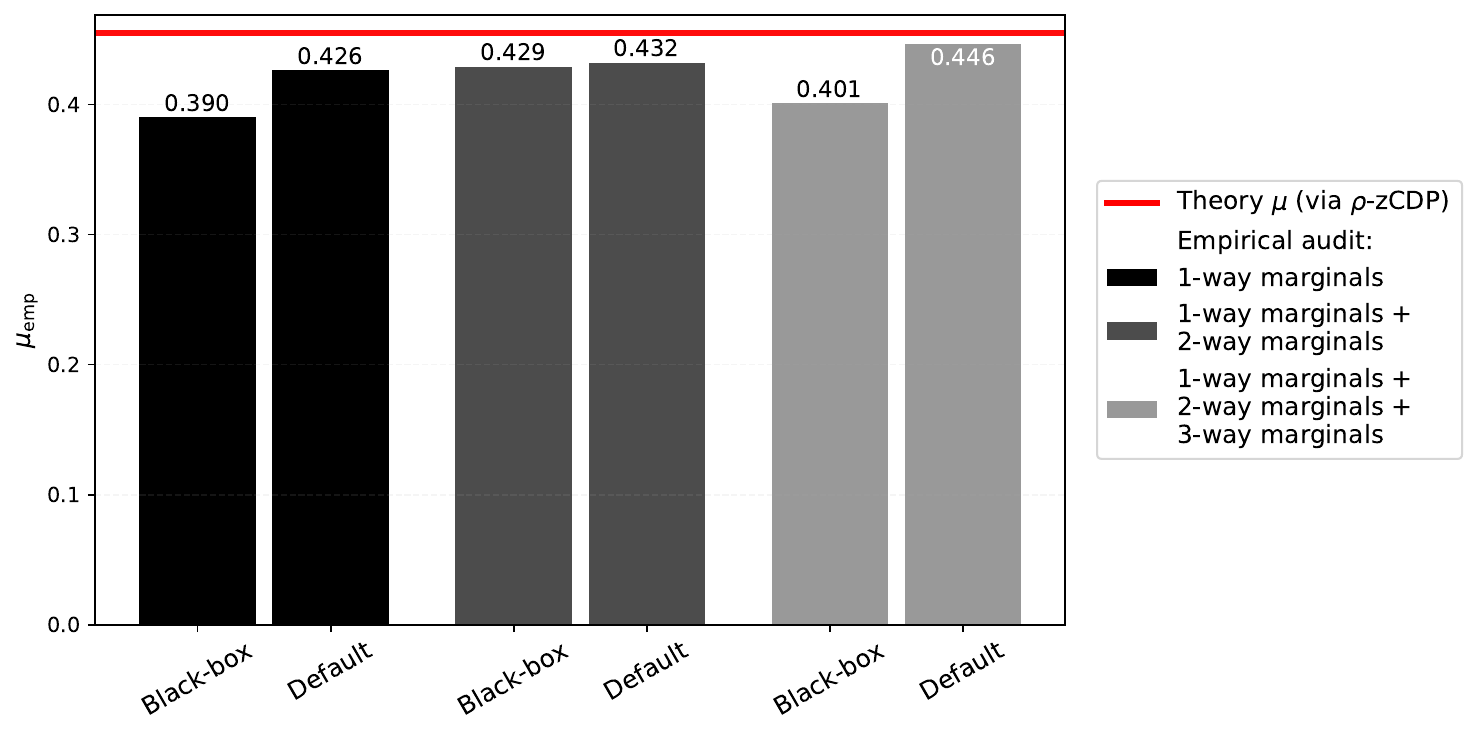}
    \vspace{-0.3cm}
	\caption{Empirical privacy $\mu_{emp}$ for MST/AIM trained with varying orders of marginals, comparing Black-box and ``Default'' (hybrid Black/White-box) threat models.
    Red line is the implied $\mu$ derived via $\rho$-zCDP.}
	\label{fig:marginals}
    \vspace{-0.5cm}
\end{figure*}

In this section, we relax the independent-marginal setting by training MST/AIM with higher-order marginals (2-way and 3-way), while keeping the dependency graph fixed.
This moves closer to the full MST/AIM pipeline while still enabling analysis within the $\mu$-GDP framework, as the privacy budget is entirely allocated to the Gaussian measurement step.

Results for both the black-box and our ``Default'' (hybrid black/white-box) threat models are shown in Fig.~\ref{fig:marginals}.
We find that incorporating higher-order marginals does not weaken the audit; on the contrary, it yields slightly tighter empirical estimates despite the increased noise per marginal due to budget splitting.
In particular, we obtain $\mu_{emp}\approx 0.44$ for 3-way marginals, compared to $\mu_{emp}\approx 0.43$ for 1-way marginals (vs. implied $\mu=0.45$).

Importantly, tight audits persist even in the black-box setting, which operates purely on synthetic data without access to internal noisy measurements.
For example, with 2-way marginals, we achieve $\mu_{emp}\approx 0.43$, closely matching the implied guarantee.
These results indicate that our findings are not specific to the independent-marginal case, but extend to more expressive marginal structures and realistic threat models.

\end{document}